\documentclass[10pt,twocolumn,final]{article}
\pdfoutput=1

\usepackage[top=2cm, left=1.5cm, bottom=2cm, right=1.5cm]{geometry}

\usepackage{latexsym}
\usepackage{graphicx}
\usepackage{balance}

\usepackage{array}
\usepackage{multirow}

\usepackage[square,numbers,sort&compress]{natbib}
\usepackage{fancyhdr}
\makeatother
\usepackage{url}
\usepackage{xcolor}
\usepackage{pdfpages}
\usepackage[title]{appendix}

\usepackage{amssymb}
\usepackage{amsmath}
\usepackage{bm}
\newcommand{\fscores}[1]{F\textsubscript{#1}-scores}
\usepackage{caption}
\newcommand{\customCaptionTable}[2]{
\vspace{-15pt}
\begin{flushleft}
\footnotesize
\textbf{Table~#1}: #2
\end{flushleft}
}

\begin{document}
\twocolumn[
  \begin{@twocolumnfalse}

\title{
Sequential Hard Mining: a data-centric approach for Mitosis Detection
}
\date{}

\vspace*{-50pt}
\begin{minipage}{\textwidth}
\centering
\author{
Maxime W. Lafarge\textsuperscript{1,2} and Viktor H. Koelzer\textsuperscript{1,2}
}
\end{minipage}

\maketitle

\vspace*{-20pt}
\begin{center}
\begin{minipage}{0.9\textwidth}
\begin{flushleft}
\footnotesize
\textsuperscript{1}\textit{Department of Biomedical Engineering, Faculty of Medicine, University of Basel, Allschwil, Switzerland} \newline
\textsuperscript{2}\textit{Institute of Medical Genetics and Pathology, University Hospital Basel, Basel, Switzerland}
\end{flushleft}
\end{minipage}
\end{center}

\vspace*{20pt}
\begin{abstract}
With a continuously growing availability of annotated datasets of mitotic figures in histology images, finding the best way to optimally use with this unprecedented amount of data to optimally train deep learning models has become a new challenge. Here, we build upon previously proposed approaches with a focus on efficient sampling of training data inspired by boosting techniques and present our candidate solutions for the two tracks of the MIDOG 2025 challenge.
\end{abstract}
\vspace*{40pt}

  \end{@twocolumnfalse}
]

%%================
\section*{Introduction}
To foster the development of high-performance mitotic figure (MF) detection algorithms and atypical mitotic figure (AMF) classifiers in the context of histopathology image analysis, the \textit{MIDOG 2025 Challenge} \cite{midog2025} has promoted the access to multiple annotated datasets and invited the mitosis detection research community to propose solutions to these two image analysis problems, by comparing the performance of candidate solutions in a controlled and independent manner.

This opportunity motivated us to propose a new approach built upon the solutions  \cite{lafarge2021midog,lafarge2022midog} we previously submitted to the MIDOG 2021 \cite{midog2021} and MIDOG 2022 challenges \cite{midog2022}, using the new annotated image datasets provided by the organizers.

With this unprecedented amount of annotated mitosis data, we employed a data-centric approach based on the sampling of training data. 
To summarize our overall approach, 1) we used a fully convolutional residual neural network (CNN) equipped with p4m-group convolution layers \cite{cohen2016groupCNN,lafarge2020roto} making the model invariant to 90-degree rotations and mirror transformations, 2) we extensively used a staining augmentation procedure on top of the standard data augmentation policies, 3) we trained models over multiple rounds and used a hard-mining procedure to sequentially update the training data sampling function.

The internal-validation and preliminary-test performance of our submitted models are reported in Table.~\ref{tab:scores}.

%================
\section*{Model Architecture}
We implemented a customized $72$-layer ResNet architecture \citep{he2016preActResNet} with a receptive field of $134{\times}134$ pixels, built upon the architecture we used in the MIDOG 2022 challenge \cite{lafarge2022midog}, to model the probability for input image pixels to be in the neighborhood of a MF/AMF.
We replaced standard convolutional layers by P4M-group convolutional layers \citep{cohen2016groupCNN} to guarantee invariance to $90$-degree rotations and mirror transformations of input images without requiring train-time or test-time rotation/mirror augmentations.

%================
\section*{Dataset Partitioning and Preparation}
To train models and evaluate their performance, we exclusively used the MIDOG++ \cite{midogpp} and MITOS-CMC  \cite{cmc2025} datasets for the MF detection task, and the MIDOG++ \cite{midogpp} and AMI-BR \cite{amibr2025} datasets for the AMF classification task.
For each task, we independently split the MIDOG++ at the image-level according to a $85$-$15$ training-validation scheme such that labels and (tumor types) were approximately balanced, the other dataset was exclusively used for training.
The provided MF and AMF annotations were used to sample positive and negative examples.
In particular, a 16-pixel spacing sampling grid was overlayed on the images of the MIDOG++ dataset to sample negative training examples.

%================
\section*{Training Procedure}
All the models were trained by minimizing the cross-entropy loss via stochastic gradient descent with momentum (initial learning rate $0.1$ and momentum $0.95$) using input mini-batches of size $16$ pixels.
Models were trained for 150000 iterations, we used a learning rate decay scheduling of factors 0.5, 0.1, 0.01 at iterations 75000, 12000 and 140000 and applied a weight decay regularization (coefficient $10^{-4}$).
Mini-batches were generated with randomly sampled image patches with a size of $166{\times}166$ pixels which produce softmax-activated probability maps of size $17{\times}17$ pixels.
All image patches were randomly transformed according to an augmentation protocol (including channel-wise intensity distortions and staining augmentation).
For evaluation purposes, we saved the weights of the model that achieved the lowest validation loss (cross-entropy) over the training period.

%================
\section*{Sequential Hard Mining}
Hard negative mining has become a standard procedure for the development of mitosis detectors since the solution proposed by \cite{cirecsan2013mitosis}, and we previously showed that this technique can be used and parameterized for a given dataset to improve classification performance by reducing the redundancy of "easily classified" examples.
Here, we broaden this concept to all misclassified training examples (not only negative training examples) as we employed an iterative approach in which every training example is associated with a sampling factor and models are sequentially trained in rounds over which training examples are sampled based on these factors.
In the first round, training examples are assigned with a sampling factor such that classes in a training batch are balanced. After training multiple models in a given round, models are compared using the validation set, and the current best performing model is used to update the sampling factor of every available training example such that correctly-classified examples are less likely to be sampled in the next round and misclassified training examples are more likely to be classified in the next round.
This procedure is repeated sequentially until no performance improvement is observed.

%================
\section*{Inference Pipeline}
For the MF detection task, once training is complete, we produced prediction maps by applying the trained model densely on test images; these maps are then binarized with a validation-based optimized cutoff value and candidate objects are identified using a connected-component extraction algorithm. The probability of the centroid pixel of the candidate objects is further compared to a validation-based F1-score-optimized cutoff value to filter final candidate detected objects.
For the AMF classification task, input patches with a center pixel above a validation-based balanced-accuracy-optimized cutoff value were considered as positive.

%================
\vspace{10pt}
\begin{table}[ht!]
\begin{center}
\caption{} %-- Virtual caption
\label{tab:scores}
\customCaptionTable{\ref{tab:scores}}{
MF detection task: Comparison of \fscores{1} of our trained models over multiple hard-mining rounds and the baseline solution of the \textit{MIDOG 2025 Challenge} on our internal validation set and the hidden preliminary test set of the challenge.
}

\newlength{\scoreLen}
\setlength{\scoreLen}{0.3\columnwidth}
\newlength{\titleLen}
\setlength{\titleLen}{0.25\columnwidth}

\newcommand{\cc}[1]{\multicolumn{1}{c}{\hspace{5pt} #1 \hspace{5pt}}}
\renewcommand{\arraystretch}{0.9}
\footnotesize
\begin{tabular}{m{\titleLen} m{\scoreLen} m{\scoreLen}}
%HEAD 1
 Hard-Mining \newline Rounds 
&  MF detection \newline Track 1 \newline (F1-score)
&  AMF classification \newline Track 2 \newline (balanced accuracy)
\\\hline\hline

\textit{Round \#1}
& \cc{$.561$}
& \cc{$.718$}
\\\hline

\textit{Round \#2}
& \cc{$.813$}
& \cc{$\mathbf{.846}$}
\\\hline

\textit{Round \#3}
& \cc{$\mathbf{.818}$}
& \cc{-}
\\\hline

\textit{preliminary test}
& \cc{-}
& \cc{$\mathbf{.812}$}
\\\hline

\textit{baseline} \cite{midog2025}
& \cc{$.767$}
& \cc{$.793$}
\\\hline

\end{tabular}
\end{center}
\end{table}
%================

%\section*{Appendix}

%%
%%*** REFERENCES
%%
\subsection*{References}
\balance
\vspace*{20pt}
\bibliographystyle{unsrtnat}
\renewcommand\refname{}
\vspace*{-40pt}
\footnotesize
\bibliography{main}

\begin{thebibliography}{12}
\providecommand{\natexlab}[1]{#1}
\providecommand{\url}[1]{\texttt{#1}}
\expandafter\ifx\csname urlstyle\endcsname\relax
  \providecommand{\doi}[1]{doi: #1}\else
  \providecommand{\doi}{doi: \begingroup \urlstyle{rm}\Url}\fi

\bibitem[Ammeling et~al.(2025)Ammeling, Aubreville, Banerjee, Bertram, Breininger, Hirling, Horvath, Stathonikos, and Veta]{midog2025}
Jonas Ammeling, Marc Aubreville, Sweta Banerjee, Christof~A. Bertram, Katharina Breininger, Dominik Hirling, Peter Horvath, Nikolas Stathonikos, and Mitko Veta.
\newblock Mitosis {Domain} {Generalization} {Challenge} 2025.
\newblock Zenodo, March 2025.
\newblock \doi{10.5281/zenodo.15077361}.
\newblock URL \url{https://doi.org/10.5281/zenodo.15077361}.

\bibitem[Lafarge and Koelzer(2021)]{lafarge2021midog}
Maxime~W Lafarge and Viktor~H Koelzer.
\newblock Rotation invariance and extensive data augmentation: A strategy for the {MI}tosis {DO}main {G}eneralization ({MIDOG}) challenge.
\newblock In \emph{International Conference on Medical Image Computing and Computer-Assisted Intervention}, 2021.

\bibitem[Lafarge and Koelzer(2022)]{lafarge2022midog}
Maxime~W Lafarge and Viktor~H Koelzer.
\newblock Fine-grained hard-negative mining: generalizing mitosis detection with a fifth of the midog 2022 dataset.
\newblock In \emph{MICCAI Challenge on Mitosis Domain Generalization}, pages 226--233. Springer, 2022.

\bibitem[Aubreville et~al.(2022)Aubreville, Stathonikos, Bertram, Klopleisch, ter Hoeve, Ciompi, Wilm, Marzahl, Donovan, Maier, et~al.]{midog2021}
Marc Aubreville, Nikolas Stathonikos, Christof~A Bertram, Robert Klopleisch, Natalie ter Hoeve, Francesco Ciompi, Frauke Wilm, Christian Marzahl, Taryn~A Donovan, Andreas Maier, et~al.
\newblock Mitosis domain generalization in histopathology images -- the {MIDOG} challenge.
\newblock \emph{Medical Image Analysis}, 2022.

\bibitem[Aubreville et~al.(2024)Aubreville, Stathonikos, Donovan, Klopfleisch, Ammeling, Ganz, Wilm, Veta, Jabari, Eckstein, et~al.]{midog2022}
Marc Aubreville, Nikolas Stathonikos, Taryn~A Donovan, Robert Klopfleisch, Jonas Ammeling, Jonathan Ganz, Frauke Wilm, Mitko Veta, Samir Jabari, Markus Eckstein, et~al.
\newblock Domain generalization across tumor types, laboratories, and species—insights from the 2022 edition of the mitosis domain generalization challenge.
\newblock \emph{Medical Image Analysis}, 94:\penalty0 103155, 2024.

\bibitem[Cohen and Welling(2016)]{cohen2016groupCNN}
Taco Cohen and Max Welling.
\newblock Group equivariant convolutional networks.
\newblock In \emph{Proceedings of the International Conference on Machine Learning (ICML)}, pages 2990--2999, 2016.

\bibitem[Lafarge et~al.(2021)Lafarge, Bekkers, Pluim, Duits, and Veta]{lafarge2020roto}
Maxime~W Lafarge, Erik~J Bekkers, Josien~PW Pluim, Remco Duits, and Mitko Veta.
\newblock Roto-translation equivariant convolutional networks: Application to histopathology image analysis.
\newblock \emph{Medical Image Analysis}, 68:\penalty0 101849, 2021.

\bibitem[He et~al.(2016)He, Zhang, Ren, and Sun]{he2016preActResNet}
Kaiming He, Xiangyu Zhang, Shaoqing Ren, and Jian Sun.
\newblock Identity mappings in deep residual networks.
\newblock In \emph{European Conference on Computer Vision (ECCV)}, pages 630--645, 2016.

\bibitem[Aubreville et~al.(2023)Aubreville, Wilm, Stathonikos, Breininger, Donovan, Jabari, Veta, Ganz, Ammeling, Van~Diest, Klopfleisch, and Bertram]{midogpp}
Marc Aubreville, Frauke Wilm, Nikolas Stathonikos, Katharina Breininger, Taryn~A. Donovan, Samir Jabari, Mitko Veta, Jonathan Ganz, Jonas Ammeling, Paul~J. Van~Diest, Robert Klopfleisch, and Christof~A. Bertram.
\newblock A comprehensive multi-domain dataset for mitotic figure detection.
\newblock \emph{Scientific Data}, 10:\penalty0 484, July 2023.

\bibitem[Aubreville et~al.(2020)Aubreville, Bertram, Donovan, Marzahl, Maier, and Klopfleisch]{cmc2025}
Marc Aubreville, Christof~A. Bertram, Taryn~A. Donovan, Christian Marzahl, Andreas Maier, and Robert Klopfleisch.
\newblock A completely annotated whole slide image dataset of canine breast cancer to aid human breast cancer research.
\newblock \emph{Scientific Data}, 7:\penalty0 417, 2020.

\bibitem[Bertram et~al.(2025)Bertram, Weiss, Donovan, Banerjee, Conrad, Ammeling, Klopfleisch, Kaltenecker, and Aubreville]{amibr2025}
Christof~A. Bertram, Viktoria Weiss, Taryn~A. Donovan, Sweta Banerjee, Thomas Conrad, Jonas Ammeling, Robert Klopfleisch, Christopher Kaltenecker, and Marc Aubreville.
\newblock Histologic {Dataset} of {Normal} and {Atypical} {Mitotic} {Figures} on {Human} {Breast} {Cancer} ({AMi}-{Br}).
\newblock pages 113--118. 2025.

\bibitem[Cire{\c{s}}an et~al.(2013)Cire{\c{s}}an, Giusti, Gambardella, and Schmidhuber]{cirecsan2013mitosis}
Dan~C Cire{\c{s}}an, Alessandro Giusti, Luca~M Gambardella, and J{\"u}rgen Schmidhuber.
\newblock Mitosis detection in breast cancer histology images with deep neural networks.
\newblock In \emph{Proceedings of the International Conference on Medical Image Computing and Computer-Assisted Intervention (MICCAI)}, 2013.

\end{thebibliography}

\newpage
\appendix

%%******************************************************************************************
%%*** *) BACKMATTER
%%%
%\newpage
%\appendix

\end{document}